\documentclass[prl,aps,twocolumn]{revtex4}
\usepackage{epsf}
\usepackage{amsmath}

\begin{document}
\title{Model of photon emission from luminescent center}

\author{Eugeniusz Chimczak}

\affiliation{Pozna\'n University of Technology, Faculty of Technical Physics, 
ul. Nieszawska 13 a, 60-965 Pozna\'n, Poland}

\date{\today}
\email{chimczak@phys.put.poznan.pl}

\begin{abstract}
Luminescence is the phenomenon investigated and applied in many disciplines of
science and technique. Spectral and kinetic measurements of luminescence provide much
information concerning the mechanism of luminescent devices. Better understanding of the
mechanism allows to improve the devices. It has been known for a long time that the spectral
and kinetic properties of luminescence depend on the temperature~\cite{curie60}. It is assumed that
fluorescence does not depend on temperature but practically lifetime of fluorescence changes
insignificantly with temperature~\cite{curie60,chimczak84}. Until now, to our knowledge, nobody describes
mathematically the dependence. Here is reported the model of photon emission permitting a
description of the dependence. It is found that lifetime of fluorescence decreases with
increasing the square root of temperature. In the model presented, the new application of
Maxwell distribution is shown.
\end{abstract}

\maketitle

Temperature influences the luminescence properties of materials. It is shown, 
in many papers, that temperature changes the shape of luminescence spectra~\cite{busse76,seo03,weman95,harukawa00,joubert87,lucas94,wolfert85,
berdowski85,suchocki87}. With increasing temperature, some bands of luminescence 
spectrum can increase or decrease in intensity. Also, in some cases, some bands 
or lines can be shifted. Temperature effects also on kinetic properties of luminescence. 
Measurements of the kinetic properties are highly important since they provide much 
information concerning the luminescent centers and the mechanism of luminescence in 
general. Therefore, numerous papers have been devoted to the lifetime of luminescence 
because the parameter determines properties of luminescence centers~\cite{chimczak84,busse76,seo03,weman95,harukawa00,joubert87,lucas94,wolfert85,
thilsing03,zhang95_2,kapoor00,berdowski85,suchocki87,leslie81,chimczak85,zhang95,bergman87,ramirez05,aizawa06}. 
In several papers there is mathematical description of the lifetime. There is mathematical 
description of dependence of the fluorescence lifetime on crystal thickness, average 
absorption coefficient and intrinsic fluorescence decay time but not on 
temperature~\cite{suchocki87}. In order to explain the temperature dependence 
of the lifetime of fluorescence, the model of photon emission from luminescent 
centers is proposed.

Let us assume that: (i) Luminescent center is some kind of trap for photons. 
(ii) Trapped photon and electron create new particle like radioactive atom. 
(iii) Motion of the particle is periodic. The period $t_{c}$ of the motion is the 
time required to complete one round trip of the motion, that is, one complete cycle. 
(iv) Speed of the particle in the motion is governed by Maxwell distribution. 
(v) Photon can only be emitted from luminescent center at the same point of the cycle 
path. (vi) During time $dt$, the particle has speeds between $v$ and $v +d v$; 
$dt=dn$ seconds, where $dn$ plays the same role as number of molecules in classical 
Maxwell distribution.

Under such assumptions the average speed of the particle is
\begin{eqnarray}
\label{eq:speed}
\overline{v}&=&\frac{a}{t_{c}} \, ,
\end{eqnarray}
where $a$ is constant. The average speed obtained from Maxwell distribution is
\begin{eqnarray}
\label{eq:speed2}
\overline{v}&=&b\sqrt{k T} \, ,
\end{eqnarray}
where $k$ is Boltzmann's constant, $T$ is the absolute temperature and $b$ is 
constant. From~(\ref{eq:speed}) and~(\ref{eq:speed2}) we obtain
\begin{eqnarray}
\label{eq:tc}
t_{c}&=&\frac{c}{\sqrt{T}} \, ,
\end{eqnarray}
where $c$ is constant. Because lifetime of luminescence $\tau$ and $t_{c}$
are constant, we can write
\begin{eqnarray}
\label{eq:tau}
\tau&=&\frac{d}{\sqrt{T}} \, ,
\end{eqnarray}
where $d$ is constant. The temperature dependence is considerably weaker 
than that observed in phosphorescence. The model proposed explains why, 
contrary to --- for example --- Cherenkov's radiation, luminescence 
appears after some time longer than about $10^{-10}$ s.

Conclusions resulted from the model are following: (i)~If the luminescent 
centers are monomolecular  and if the rectangular exciting pulse is very 
long then, during the time $t=t_{c}$ after the end of the exciting pulse, 
the luminescence intensity is constant and equal to luminescence intensity 
at the end of the pulse, when direct impact excitation is the only process. 
Time dependence of the luminescence after the end of the pulse is staircase 
curve. (ii)~Trapped photon can only be emitted from luminescent center at time 
$t=n t_{c}$, where $n$ is natural number.

This work was supported by Pozna\'n University of Technology Research 
Program 62-208/07 - BW.

\end{document}